\documentclass[prb,twocolumn,amsmath,amssymb,superscriptaddress,showpacs]{revtex4}

\usepackage{bm}
\usepackage{graphicx}
\usepackage{multirow}
\usepackage{color}

\begin{document}

\title{Optical selection rules for excitonic Rydberg series in the massive Dirac cones of hexagonal 2D materials}

\author{Pu Gong}
\affiliation{Department of Physics, and Center for Theoretical and Computational Physics, The University of Hong Kong, Hong Kong, China}

\author{Hongyi Yu} \thanks{yuhongyi@hku.hk}
\affiliation{Department of Physics, and Center for Theoretical and Computational Physics, The University of Hong Kong, Hong Kong,  China}

\author{Yong Wang}
\affiliation{School of Physics, Nankai University, Tianjin 300071, China}
\affiliation{Department of Physics, and Center for Theoretical and Computational Physics, The University of Hong Kong, Hong Kong,  China}

\author{Wang Yao}
\affiliation{Department of Physics, and Center for Theoretical and Computational Physics, The University of Hong Kong, Hong Kong,  China}

\begin{abstract}

We investigate the optical transition selection rules for excitonic Rydberg series formed in massive Dirac cones. The entanglement of the exciton envelop function with the pseudospin texture leads to anomalous selection rules for one-photon generation of excitons, where $d$-orbitals can be excited with the opposite helicity selection rule from the $s$-orbitals in a given valley. The trigonal warping effects in realistic hexagonal lattices further renders more excited states bright, where $p$-orbitals can also be accessed by one-photon excitation with the opposite valley selection rules to the $s$-orbitals. The one-photon generation of exciton in the various states and the intra-excitonic transition between these states are both dictated by the discrete in-plane rotational symmetry of the lattices, and our results show that in hexagonal 2D materials the symmetry allowed transitions are enabled when trigonal warping effects are included in the massive Dirac fermion model. In monolayer transition metal dichalcogenides where excitons can be generated by visible light and intra-excitonic transitions can be induced by infrared light, we give the strength of these optical transitions, estimated using modified hydrogen-like envelope functions combined with the optical transition matrix elements between the Bloch states calculated at various $k$ points.

 \end{abstract}

 \date{\today}


 \maketitle

\section{Introduction}

The study of two-dimensional (2D) Dirac materials has been one of the most active fields of research today. Seminal examples are graphene and surface states of topological insulators where the massless Dirac cones give rise to exotic properties of wide scientific and technological interest.~\cite{Rise of Graphene, RMP TI, RMP XL Qi} The Dirac cones become massive ones when an energy gap is opened at the Dirac points, which can be introduced by inversion symmetry breaking in graphene,~\cite{Xiao and Yao PRL07, Yao Valleyoptics 08} or by the tunneling between the top and bottom surfaces in a topological insulator thin film.~\cite{Hz MDF TI PRB10}
Studies based on these model systems have discovered interesting properties of the massive Dirac cones such as the valley selection rules of the interband optical transitions transitions.~\cite{Yao Valleyoptics 08}
The emergence of monolayer group-VIB transition-metal dichalcogenides (TMDs) has offered a practical platform to explore such optical properties of massive Dirac fermions.~\cite{QH Wang nnano rev12, nphys TMDs rev14, Guibin rev15}
These compounds are of the chemical composition of $MX_2$ ($M=$ W or Mo, $X=$ S or Se). Monolayer TMDs are $X$-$M$-$X$ covalently bonded 2D hexagonal lattices, and their stacking and bounding by the weak van der Waals interaction form the layered bulk crystals. Thinning down from bulk to monolayers, a most remarkable change is the crossover from indirect to a direct band gap semiconductor.~\cite{Mak MoS2 PRL2010, Feng Wang nanolett2010, Direct Gap nnano tech2014}
In the monolayer TMDs, both the conduction and valence band edges are at the degenerate but inequivalent K and -K corners of the hexagonal Brillouin zone, and interestingly the minimal $\mathbf{k \cdot p}$ Hamiltonian describing these band edges is a valley dependent massive Dirac fermion model.~\cite{Yao Coupled Spin Valley 2012}

With a direct gap in the visible frequency range, monolayer TMDs are ideal for the exploration of optical properties of massive Dirac fermions.
Optical field can excite interband transitions, creating an electron in the upper branch and a hole in the lower branch of the massive Dirac cones.
The attractive Coulomb interaction will bind the optically excited electron-hole pair into a hydrogen-like state, known as exciton, a most fundamental optical excitation found in semiconductors. Remarkably, in monolayer TMDs, Coulomb interaction is particularly strong due to the 2D geometry as well as the large effective mass of the Dirac cone, resulting in tightly bound excitons with large binding energies.  Optical resonances corresponding to excitons and their charged counterpart (i.e. binding an extra electron or hole) have been observed in monolayer TMDs.~\cite{Valley Coherence nnano2013, Electrical Control excitons ncomm2013, Mak Trions nmater2013}
The large spectral separation between the charged and neutral excitons point to an exceptionally large binding energy in the order of hundreds of meV,~\cite{Valley Coherence nnano2013, Electrical Control excitons ncomm2013, Mak Trions nmater2013} consistent with first principles calculations,~\cite{J. Feng nphoton2012, S. Louie MoS2 specturm calculation PRL2013,  Quaisi band calculation 2012, Excitonic effects calculation 2012, Binding energy from first principles 2012, Quasiparticle optical properties strained calculation 2013}
and experiments where the binding energies are extracted from measuring the excitonic excited states
~\cite{Heinz binding energy PRL2014, Jie Shan binding energy PRL2014, X Zhang S Louie TPA  2014, XD Cui TPA 2014, Urbaszek TPA and SHG PRL2015}
or the quasi-particle (electronic) band gap.~\cite{Shih MoS2 2.15eV, Berkeley bandgap renormalization nmat2014, S. Louie MoS2 specturm calculation PRL2013, Heinz binding energy PRL2014, Jie Shan binding energy PRL2014}

The formations of tightly bound excitons in the valley dependent massive Dirac cones are of particular interest. With a Wannier type wavefunction,~\cite{S. Louie MoS2 specturm calculation PRL2013} the excitonic ground states inherit the valley optical selection rules of the band to band transition,~\cite{Yao Coupled Spin Valley 2012, Yao Valleyoptics 08}
allowing their valley specific interconversion with photons of selected helicity.
Based on the selection rules, optical generation of exciton valley polarizations and valley coherence have been demonstrated in monolayer TMDs,~\cite{Valley Coherence nnano2013, Mak Valley polarization nnano2012, Zeng Valley polarization nnano2012, CaoTing Valley physics ncomm2012} and optical studies of various valley related phenomena became possible.~\cite{valley dynamics DSun 2013, Valley lifetime measurement 2014, VHE transisitor 2014, Spin-valley photocurrent 2015} Compared to the excitons formed in conventional semiconductors like GaAs, the underlying pseudospin texture of the massive Dirac cones can play an important role here. In principle the envelope function of the exciton is entangled with the pseudospin texture lying in the periodic part of the electron and hole Bloch functions, which need to be properly accounted beyond the envelope function approximation.
For example, certain gauge choice of the pseudospin wavefunction can lead to a $p$-like phase winding in the exciton ground state envelope function.~\cite{Koch p-state brightness 2015} This artificial phase winding goes away under the proper gauge choice where the conventional rotational symmetry of the envelope functions can be restored. Such a gauge choice is used hereafter, detail definition follows.~\cite{Di Xiao Berry phase and exciton 2015}
The gauge field (Berry curvature) in the massive Dirac cone can also modify the Coulomb interaction between the electron and hole,
lifting the degeneracy between exciton states with the same angular momentum quantum number but different magnetic quantum numbers.~\cite{Excitons in Bilayer Graphene nano2010, Excitons in TI with magnetic gap PRB2011, Di Xiao Berry phase and exciton 2015, Imamoglu p-state Berry splitting 2015}

In this paper, we investigate the optical transition selection rules for excitonic Rydberg series formed in massive Dirac cones. We find that the entanglement of the exciton envelope function with the pseudospin texture leads to anomalous selection rules for the one-photon generation of excitons, which cannot be accounted in the envelope function approximation. Besides those $s$-like orbitals, the bright excitonic states also include the $d$-orbitals which have the opposite valley dependent helicity selection rule from the $s$-orbitals. This is in contrast to excitons in the envelope function approximation where only the $s$-orbitals are optically bright. The transition strength to the $d$-states can be enhanced by the effective mass correction ($k^2 \sigma_z$ term) to the massive Dirac cones. Moreover, we find nontrivial modifications of the transition selection rules by the inevitable trigonal warping effects in realistic hexagonal lattices, where the reduction of in-plane rotational symmetry renders more excited states bright, including the $p$-orbitals which also have the opposite valley selection rules to the $s$-orbitals. We further analyze the one-photon transitions between the excitonic states, which becomes highly relevant in monolayer TMDs with the large energy separations of the Rydberg series. In massive Dirac cones with the trigonal warping effects, the intra-excitonic transitions not possible in the envelope function approximation can now be accessed as long as they are allowed by the three-fold in-plane rotational symmetry.

In monolayer TMDs, we have estimated the strength of the one-photon generation of excitonic states, and the transition strength between the excitonic states, by using the first principle values of envelope function $\mathbf k$-space spreading combined with the optical transition matrix elements between the Bloch states calculated at various $k$ points. The transition strength of the 2$p$-orbitals is found to be just one order smaller compared to that of the 1$s$ ground states. In the calculation of the band-to-band transition, we have used respectively first principle calculations, a two-band $\mathbf k\cdot\mathbf p$ model and a three-band tight-binding model. Through comparing these different approaches, we find that the $\mathbf k\cdot\mathbf p$ model and the tight-binding model can not accurately give the strength of the excitonic transitions, even when other quantities including the band dispersions and Berry curvatures are well captured in the models.

The rest of the paper is organized as follows. In section \ref{sectionMDF}, we start with the massive Dirac fermion model with the continuous in-plane rotational symmetry and show the anomalous valley selection rules for the $d$-orbital excitonic states. In secion \ref{secionTW}, quadratic $k$ terms are added to the massive Dirac fermion model to account for the trigonal warping effects and the effective mass corrections to the electron and hole, and the anomalous valley selective optical transitions to the $p$-orbital excitonic states are derived. In section \ref{sectionIntra}, the intra-excitonic transitions between the excitonic states are discussed. In section \ref{sectionTMDs}, we give quantitative estimations for the strength of the various excitonic transitions in monolayer TMDs.

\section{Excitonic transition selection rules in massive Dirac fermion model \label{sectionMDF}}

To begin with, we employ the massive Dirac fermion (MDF) Hamiltonian
\begin{equation}
\hat{H}_0 = at(\tau  k_x\hat{\sigma}_x + k_y\hat{\sigma}_y) + \frac{\Delta}{2} \hat{\sigma}_z, \label{MDF}
\end{equation}
where $\mathbf{k}=(k_{x},k_{y})$ is the wave vector relative to $\tau$K point, $\tau = \pm 1$ is the index of the two ($\pm$K) valleys that are related by time reversal symmetry, $\hat{\sigma}_{x/y/z}$ denote the Pauli matrices for pseudospin spanning the conduction and valence states at $\tau$K, $a$ the lattice constant, $t$ the effective hopping integral, and $\Delta$ the energy gap at $\pm$K. 

Eq. (\ref{MDF}) is the minimum Hamiltonian describing the conduction and valence band edges in monolayer TMDs and in graphene with a staggered sublattice potential.~\cite{Xiao and Yao PRL07, Yao Coupled Spin Valley 2012}
In graphene, the pseudospin Hilbert space is spanned by the two carbon $p_z$-orbitals respectively on the A and B sites in a unit cell, while in TMDs, the pseudospin Hilbert space is spanned by the two $d$-orbitals of the metal atom: $\left|\phi_{c}\right\rangle =\left|d_{z^{2}}\right\rangle$,
$\left|\phi_{v}^{\tau}\right\rangle =\frac{1}{\sqrt{2}}(\left|d_{x^{2}-y^{2}}\right\rangle +i\tau\left|d_{xy}\right\rangle )$.
Spin-orbit coupling (SOC) is not considered here. This is because  graphene and monolayer TMDs both have a mirror symmetry in the out-of-plane (z) direction, which dictates that the spin-orbit coupling can only involve the spin component in the z-direction. Optical transition under normal incidence conserves the spin component in the z-direction. Therefore, spin-orbit coupling of the above form will only affect the transition energy but not the transition strength. 

In the rigorous sense, an exciton consisting of two massive Dirac fermions (Eq. (\ref{MDF})) with zero center-of-mass momentum is described by the Dirac Coulomb Hamiltonian~\cite{TMD MDF Coulomb PRB13, Koch p-state brightness 2015}
\begin{align}
\left(\Delta\hat{\sigma}_{z}+\frac{2at}{\hbar}\hat{\boldsymbol{\sigma}}\cdot\mathbf{p}-V(\mathbf{r})\right)\Psi_{X}(\mathbf{r})=E_{X}\Psi_{X}(\mathbf{r}).
\label{ExcitonEquation}
\end{align} 
Here we assume both the electron and hole are in the valley $\tau=+1$. $\mathbf{r}\equiv(x,y)=\left(r\cos\phi,r\sin\phi\right)$ is the electron-hole relative position, $\mathbf{p}\equiv-i\hbar\left(\frac{\partial}{\partial{x}},\frac{\partial}{\partial{y}}\right)$ is the momentum operator of relative motion, $\hat{\boldsymbol{\sigma}}=(\hat{\sigma}_x,\hat{\sigma}_y)$ is the Pauli matrices for pseudospin, $-V(\mathbf{r})$ is the Coulomb attraction between the electron and hole. 

The problem for $V(\mathbf{r})=\frac{e^{2}}{\epsilon r}$ with a homogeneous dielectric constant $\epsilon$ has been throughly investigated by earlier studies, here the results are directly quoted~\cite{TMD MDF Coulomb PRB13, Koch p-state brightness 2015}. The exciton wave function in general is a two-component spinor $\Psi_{X}(\mathbf{r})=\left(e^{i\left(j-\frac{1}{2}\right)\phi}\rho_{nj}\left(r\right),e^{i\left(j+\frac{1}{2}\right)\phi}\rho'_{nj}\left(r\right)\right)^{T}$, with its energy given by $E_{nj}=\Delta\frac{n+\sqrt{j^{2}-\alpha^{2}/4}}{\sqrt{\alpha^{2}/4+\left(n+\sqrt{j^{2}-\alpha^{2}/4}\right)^{2}}}$. Here $n=0,1,...$ is the principal quantum number, and $j=\pm\frac{1}{2},\pm\frac{3}{2},...$ the quantum number of total angular momentum $\hat{L}_{z}+\frac{\hat{\sigma}_{z}}{2}$ with $\hat{L}_{z}\equiv\left(\mathbf{r}\times\mathbf{p}\right)_{z}$ the out-of-plane component of the angular momentum. $\alpha\equiv e^{2}/\epsilon at$ is the effective fine structure constant which characterizes the pseudospin-orbit coupling strength of the exciton. Since the minimum $|j|$ value is $1/2$, $\alpha^2>1$ then corresponds to a strong coupling regime which leads to the so-called excitonic collapse.~\cite{TMD MDF Coulomb PRB13, Koch p-state brightness 2015} While in the weak coupling limit $\alpha^{2}\ll 1$, the solution returns to the Rydberg series of non-relativistic 2D hydrogen model, and the exciton wave function will be reduced to a scalar form $\Psi_{X}(\mathbf{r})=e^{i\left(j-\frac{1}{2}\right)\phi}\rho_{nj}\left(r\right)$. We note that the effective fine structure constant can also be written as $\alpha\equiv\frac{e^{2}}{\epsilon at}=\frac{2at}{\Delta a_{B}}$ in the week coupling limit, with $a_{B}=\frac{\hbar^{2}\epsilon}{2\mu e^{2}}=\frac{2\epsilon a^{2}t^{2}}{\Delta e^{2}}$ the Bohr radius of the lowest energy (1s) exciton and $\mu\equiv\frac{\hbar^{2}\Delta}{4a^{2}t^{2}}$ the reduced mass. 

In monolayer TMDs, due to its small but finite thickness and the finite background screening, the effective dielectric constant $\epsilon=\epsilon(r)$ has a nonlocal character which depends strongly on the electron-hole relative distance $r$.\cite{Louie dielectric dependence PRB16} Thus an effective fine structure constant obtained using $\alpha\equiv\frac{e^{2}}{\epsilon at}$ becomes problematic as there is no well defined homogeneous dielectric constant $\epsilon$. On the other hand, the Bohr radius $a_{B}$ in monolayer TMDs is found to be in the range from $1$ nm to several nm according to first principle calculations and theoretical analysis. If monolayer TMDs are in the weak pseudospin-orbit coupling regime, then the effective fine structure constant can be obtained as $\alpha=\frac{2at}{\Delta a_{B}}$ given that the weak coupling criterion $\alpha^{2}\ll 1$ is satisfied. Taking $a_{B}\sim1$ nm together with the typical parameter values ($a\sim3.2$~\AA, $t\sim1.2$ eV and $\Delta\sim2$ eV), we find $\alpha^{2}=\left(\frac{2at}{\Delta a_{B}}\right)^{2}\sim0.2$ which is indeed much smaller than $1$. 

From the above analysis, the monolayer TMD system shall be treated in weak pseudospin-orbit coupling regime with a scalar wave function $\Psi_{X}(\mathbf{r})$. Keeping up to the second order ($\nabla^{2}$) terms, the effective Hamiltonian $\hat{H}_\textrm{eff}$ for $\Psi_{X}(\mathbf{r})$ has a perturbative form~\cite{Di Xiao Berry phase and exciton 2015}
\begin{align}
\hat{H}_\textrm{eff}=~&\Delta-\frac{\hbar^{2}}{2\mu}\nabla^{2}-V(\mathbf{r}) \nonumber\\
&-\frac{a^{2}t^{2}}{2\Delta^{2}}\left(\left(\nabla V\times i\nabla\right)_{z}+\frac{1}{2}\nabla^{2}V\right). \label{Wannier rspace}
\end{align}
Here the second line comes from the pseudospin-orbit coupling. As $V(\mathbf{r})=V(r)$ is a central-force potential, it is not difficult to verify that 
\begin{align}
\left[\hat{L}_{z}, \hat{H}_\textrm{eff}\right]=\left[\exp\left(-\frac{i}{\hbar}\hat{L}_{z}\phi\right), \hat{H}_\textrm{eff}\right]=0. \label{LzCommu}
\end{align} 
Thus in the weak pseudospin-orbit coupling regime $\Psi_{X}(\mathbf{r})$ is rotational symmetric with a form $\Psi_{X}(\mathbf{r})=\frac{\sqrt{S}}{2\pi}e^{im\phi}\rho_{nm}(r)$, where $m=0, \pm1, \pm2,...$ is the angular quantum number of $\hat{L}_{z}$.

In monolayer TMDs the eigenstates of the MDF Hamiltonian Eq. (\ref{MDF}) correspond to the periodic part $u_{c/v,\mathbf{k}}(\mathbf r)$ of the Bloch function, with $\phi_{c/v,\mathbf{k}}(\mathbf r)=e^{i(\mathbf K+\mathbf k)\cdot\mathbf r}u_{c/v,\mathbf{k}}(\mathbf r)$ the Bloch states. Taking into account the lattice structure, the full form of the exciton wave function is $\Psi_{X}(\mathbf{r}_{e},\mathbf{r}_{h})=\sum_{\mathbf{k}}\Phi_{X}(\mathbf{k})\phi_{c,\mathbf{k}}(\mathbf{r}_{e})\phi^*_{v,\mathbf{k}}(\mathbf{r}_{h})$, where the envelope function $\Phi_{X}(\mathbf{k})$ is connected to $\Psi_{X}(\mathbf{r})$ by a Fourier transform $\Psi_{X}(\mathbf{r})=\sum_{\mathbf{k}}\Phi_{X}(\mathbf{k})e^{i\mathbf{k}\cdot\mathbf{r}}$. So $\Phi_{X}(\mathbf{k})=\frac{2\pi}{\sqrt{S}}e^{im\theta}\rho_{nm}(k)$ is rotational symmetric as $\Psi_{X}(\mathbf{r})$. As pointed out in Ref. [\onlinecite{Di Xiao Berry phase and exciton 2015, Imamoglu p-state Berry splitting 2015}], $\Phi_{X}(\mathbf{k})$ satisfies the Wannier equation with the Berry curvature corrections, whose form can be derived from $\hat{H}_\textrm{eff}\Psi_{X}(\mathbf{r})=E_X\Psi_{X}(\mathbf{r})$.

In the region $k\simeq a_{B}^{-1}$ where $\Phi_{X}(\mathbf{k})$ mainly distributes, there is $\left(\frac{2atk}{\Delta}\right)^{2}\lesssim\left(\frac{2at}{\Delta a_{B}}\right)^{2}\ll1$. Thus we can do a perturbation expansion on $\frac{atk}{\Delta}$ up to the second order that
\begin{subequations}
\label{MDFeigenstate}
\begin{align}
&\left|u_{c,\mathbf{k}}\right\rangle =\left(1-\frac{a^{2}t^{2}k^{2}}{2\Delta^{2}}\right)\left|u_{c,0}\right\rangle +\frac{at}{\Delta} ke^{i\theta}\left|u_{v,0}\right\rangle, \label{MDFcstate} \\
&\left|u_{v,\mathbf{k}}\right\rangle =\left(1-\frac{a^{2}t^{2}k^{2}}{2\Delta^{2}}\right)\left|u_{v,0}\right\rangle -\frac{at}{\Delta}ke^{-i\theta}\left|u_{c,0}\right\rangle. \label{MDFvstate}
\end{align}
\end{subequations}
Note that the weak but finite pseudospin-orbit coupling in the MDF Hamiltonian (Eq. (\ref{MDF})) leads to a $\mathbf k$-dependent pseudospin texture characterized by a finite Berry curvature $\Omega=a^2t^2/\Delta^2$ in the above expansion. Recently two parallel works have investigated how the Berry curvature can manifest in the exciton formation, by affecting the intraband dynamics of the electron and hole. The result is an angular momentum dependent energy shift to the excitons \cite{Di Xiao Berry phase and exciton 2015, Imamoglu p-state Berry splitting 2015}, which originates from the modification of the intraband direct Coulomb interaction by the pseudospin texture. 

Here in this paper, we focus on how the pseudospin texture affects an interband process, i.e. the coupling between the exciton and the light field. At K valley ($\tau=+1$), band-to-band transitions at a given $\mathbf{k}$-point induced by $\sigma \pm$ light is described by the optical transition matrix elements: $p_{\pm}(\mathbf{k})\equiv p_{x}(\mathbf{k})\pm ip_{y}(\mathbf{k})$
where  $\mathbf p$ are the interband matrix elements of momentum operator: $\mathbf{p}(\mathbf{k})=\frac{m_0}{\hbar}\langle u_{c,\mathbf k}|\nabla_{\mathbf k}\hat{H}_0|u_{v,\mathbf k}\rangle$. The transition matrix elements at the two valleys $p_{\pm}^{\tau}(\mathbf{k})$ are related through a time reversal relation $p_{\pm}^{-}(\mathbf{k})=-(p_{\mp}^{+}(-\mathbf{k}))^{*}$, hereafter we focus on the K valley and drop the $\tau$ superscript. At K valley we have
\begin{subequations}
\label{MDFpcv}
\begin{align}
&p_{+}(\mathbf{k})=\left(1-\frac{a^{2}t^{2}k^{2}}{\Delta^{2}}\right)p_0, \label{MDFp+} \\
&p_{-}(\mathbf{k})=-\left(\frac{a^{2}t^{2}k^{2}}{\Delta^{2}}\right)e^{-2i\theta}p_0.
\label{MDFp-}
\end{align}
\end{subequations}
Here $p_0\equiv 2m_{0}at/\hbar$. The effect of the pseudospin texture appears as the $\mathbf k$-dependent terms in the above optical transition matrix elements.

In the envelope function approximation the electron/hole Bloch function is approximated by $\phi_{c/v,\mathbf{k}}(\mathbf r)=e^{i(\mathbf K+\mathbf k)\cdot\mathbf r}u_{c/v,\mathbf{k}}(\mathbf r)\approx e^{i(\mathbf K+\mathbf k)\cdot\mathbf r}u_{c/v,0}(\mathbf r)$, an exciton wavefunction is then given by the direct product of the envelope function and the periodic part of the Bloch function at the high symmetry K point. Therefore, the excitonic selection rules are determined by the integral of the envelope function ($s$-orbital being the only bright one), together with the selection rule of band-to-band transition at the high symmetry point. At K point we have $p_{+}=p_0$ and $p_{-}=0$, i.e. the inter-band transition can only be excited by $\sigma +$ light.~\cite{Yao Coupled Spin Valley 2012}
However, if we go beyond the envelope function approximation, more bright states with distinct selection rules arise.

With the $\mathbf k$-dependence of the periodic part of the Bloch function (c.f. Eq. (\ref{MDFvstate}) ), the envelope function gets entangled with the pseudospin texture. Since the exciton wavefunction is a linear superposition of electron-hole pairs $\phi_{c,\mathbf{k}}\phi^*_{v,\mathbf{k}}$ at the various $\mathbf{k}$-points, its optical transition matrix element is  the coherent superposition of the band-to-band transition at these $\mathbf{k}$-points.
At a finite $\mathbf k$, the band-to-band transition has finite coupling to both $\sigma+$ and $\sigma-$ light, so the polarization selection rule is in fact an elliptical one, with the ellipticity depending on $\mathbf k$ and major axis of the elliptical polarization depending on $\theta$ (Fig. 1(a)). The latter dependence gives rise to the anomalous excitonic selection rules beyond the envelope function approximation.

As we discussed, in the weak coupling regime the envelope function for exciton can be expressed as $\Phi_{X}(\mathbf{k})=\frac{2\pi}{\sqrt{S}}\rho_{nm}(k)e^{im\theta}$, the coupling of the excitonic transition  at K valley to $\sigma \pm$ polarized light is characterized by the optical transition matrix element
\begin{align}
\left\langle \Psi_{X}\right|\mathbf{\epsilon}_{\pm}\cdot\hat{\mathbf{p}}\left|0\right\rangle =\frac{\sqrt{S}}{2\pi}\int\rho_{nm}^{*}(k)e^{-im\theta}p_{\pm}(\mathbf{k})d\mathbf{k} \label{matrixelement}
\end{align}
where $\mathbf{\epsilon}_{\pm}=\frac{1}{\sqrt{2}}(\mathbf{x}\pm i\mathbf{y})$ is the unit vector for $\sigma \pm$ polarization and we have used $\sum_{\mathbf{k}}=\frac{S}{(2\pi)^{2}}\int d\mathbf{k}$ where $S$ is the area of the box normalization \cite{BoxNorm}. We find
\begin{subequations}
\label{selectionMDF}
\begin{align}
&\left\langle\Psi_{X}\right|\mathbf{\epsilon}_{+}\cdot\hat{\mathbf{p}}\left|0\right\rangle=I_{s}\delta_{m,0}, \label{MDFplus}\\
&\left\langle\Psi_{X}\right|\mathbf{\epsilon}_{-}\cdot\hat{\mathbf{p}}\left|0\right\rangle=I_{d}\delta_{m,-2}, \label{MDFminus}
 \end{align}
\end{subequations}
where the kronecker deltas arise from the integration over the angular variable $\theta$, and
\begin{align}
&I_{s} \sim \sqrt{S}p_0\int_{0}^{2a_{B}^{-1}}
 \rho_{n,0}^{*}(k)\left(1-\frac{a^{2}t^{2}k^{2}}{\Delta^{2}}\right)kdk, \notag\\
&I_{d}  \sim -\sqrt{S}p_0\frac{a^{2}t^{2}}{\Delta^{2}}\int_{0}^{2a_{B}^{-1}}
 \rho_{n,-2}^{*}(k)k^{3}dk. \label{dipolestrengthMDF}
\end{align}
Since $\rho_{nm}(k)$ decays fast with $k$ for $k>a_B^{-1}$, in the above we have restricted the integral range to $[0,2a_B^{-1}]$ in which our perturbation treatment (Eq. (\ref{MDFcstate}) and (\ref{MDFvstate})) is well justified. Such an integral range is found to give the correct order of magnitude to the transition strength compared to that using rigorous solutions of $p_\pm(\mathbf{k})$. Eq. (\ref{MDFplus}) and (\ref{MDFminus}) give the transition selection rules for one-photon generation of exciton states at K valley: the $s$ state exciton ($m=0$) is excited by $\sigma +$ polarized  light, while the $d_{-}$ state ($m=-2$) is excited by $\sigma -$ polarized one. The $\sigma-$ optical selection rule for the $d_-$ state is a direct consequence of Eq. (\ref{MDFp-}) from the pseudospin texture. It also slightly reduces the optical transition matrix element $I_s$. Taking the time reversal of Eq. (\ref{MDFplus}) and (\ref{MDFminus}), we find the selection rules at -K valley where $\sigma -$ light generates $s$ state exciton and $\sigma +$ excite $d_{+}$ state.

\begin{figure}
\includegraphics[width=8cm]{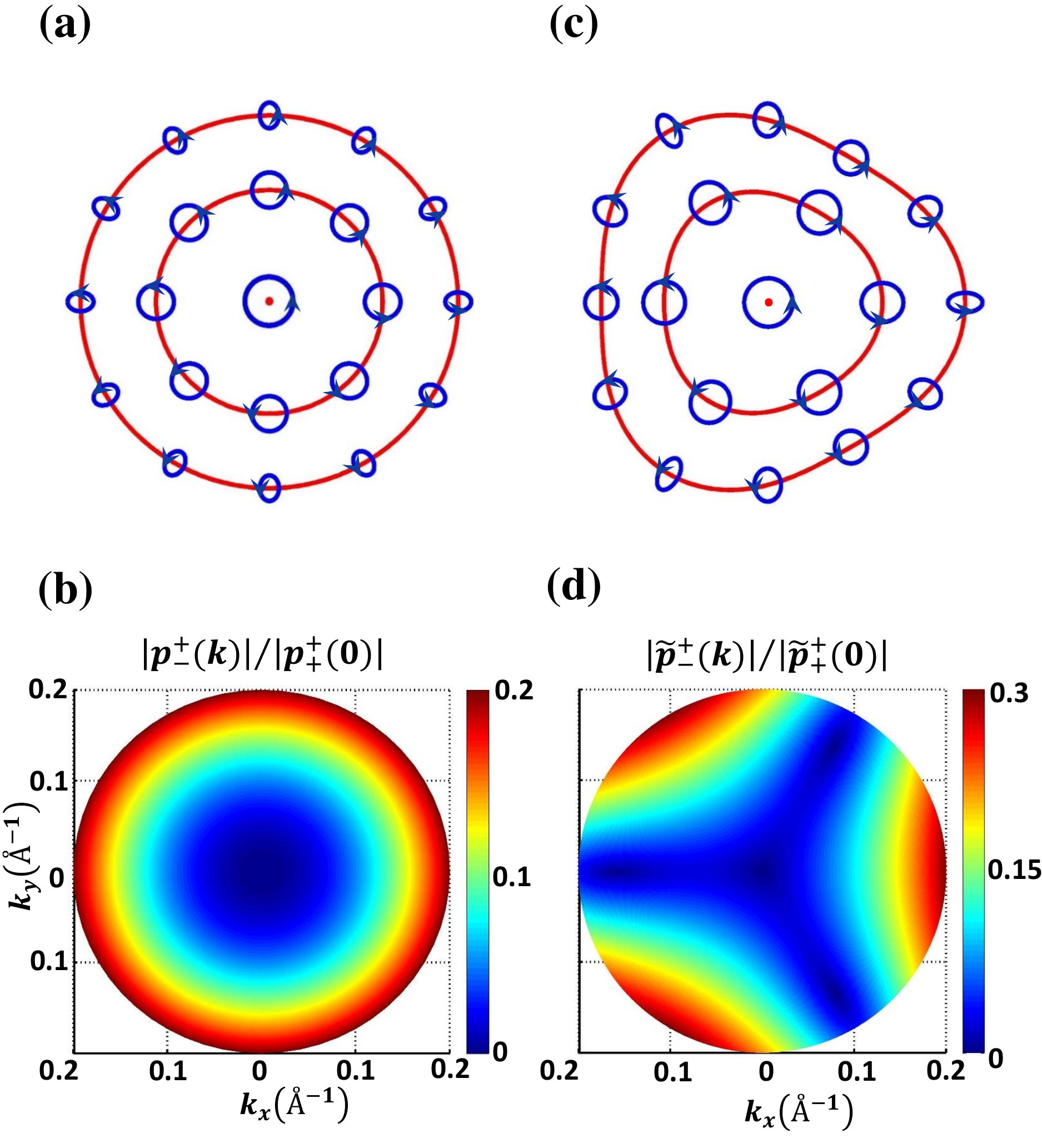}
\caption{(a) The polarization of the band-to-band transitions at the various $\mathbf{k}$-points in the K valley, calculated using the massive Dirac fermion model (Eq.~(\ref{MDF})). The polarization is in general elliptical, as shown by the blue elliptical circles with the arrows indicating the helicity. The ellipticity increase with $k$. At $\mathbf k=0$ the selection rule reduces to circular ($\sigma+$). Decomposing the elliptical polarization in the circularly polarized basis, the $\sigma+$ circularly polarized one is the dominating component. The minor $\sigma-$ polarized component is shown in (b), where magnitude of the optical transition matrix element $p_{-}(\mathbf{k})$ is plotted in the unit $|p_{+}(0)|$.
(c,d) The polarization calculated with the massive Dirac fermion model including trigonal warping effects (Eq.~(\ref{2band TW H})), with the parameters from Ref.~\onlinecite{Fal'ko kp review 2015}.}
\end{figure}

\section{Trigonal warping effects on the selection rules \label{secionTW}}

Eq. (\ref{MDFplus}) and (\ref{MDFminus}) are the consequences of the continuous in-plane rotational symmetry of the Hamiltonian in Eq. (\ref{MDF}). In reality, either graphene or monolayer TMDs lattice only has the discrete 3-fold rotational symmetry. As a result, the dispersion about the $\pm$K points is not spherical, but has trigonal warping which is more pronounced at larger $\mathbf k$. This is reflected in the $\mathbf k\cdot\mathbf p$ Hamiltonian by including terms second order in $\mathbf{k}$~\cite{Fal'ko TW 2-band 2013},
\begin{eqnarray}
\tilde{H}_{K}(\mathbf{k})&=& at( k_+\hat{\sigma}_- + k_-\hat{\sigma}_+) + \frac{\Delta}{2} \hat{\sigma}_z + \frac{B}{2} k^2 + \frac{D}{2} \sigma_z k^2  \notag \\
&& - \kappa ( k_+^2 \hat{\sigma}_+ + k_-^2 \hat{\sigma}_-)  \label{2band TW H}
 \end{eqnarray}
where $k_{\pm}=k_{x}\pm i k_{y}$. $B$ and $D$ terms account for the effective mass corrections to the hole and electron, which preserves the continuous rotational symmetry. The $B$ term causes the effective mass difference between the electron and hole, while the $D$ term can modify the transition strength to the exciton $d$ state as shown below. $\kappa$ term is another second order correction that characterizes the degree of trigonal warping effect, which reduces the rotational symmetry to a discrete three-fold one. The eigenstates from perturbation expansion are
\begin{subequations}\label{TWeigenstates}
\begin{align}
&\left|\tilde{u}_{c,\mathbf{k}}\right\rangle =\left(1-\frac{a^{2}t^{2}k^{2}}{2\Delta^{2}}\right)\left|u_{c,0}\right\rangle +\left(\frac{at}{\Delta}k_{+}-\frac{\kappa}{\Delta}k_{-}^{2}\right)\left|u_{v,0}\right\rangle, \label{TWcstate} \\
&\left|\tilde{u}_{v,\mathbf{k}}\right\rangle =-\left(\frac{at}{\Delta}k_{-}-\frac{\kappa}{\Delta}k_{+}^{2}\right)\left|u_{c,0}\right\rangle +\left(1-\frac{a^{2}t^{2}k^{2}}{2\Delta^{2}}\right)\left|u_{v,0}\right\rangle. \label{TWvstate}
\end{align}
\end{subequations}

The optical matrix elements for the band-to-band transitions are
\begin{subequations}\label{pTW}
\begin{align}
&\tilde{p}_{+}(\mathbf{k})=
p_0\left[1-(\frac{a^{2}t^{2}}{\Delta^{2}}+\frac{D}{\Delta})k^{2}\right],
\label{PplusTW}\\
&\tilde{p}_{-}(\mathbf{k})= -p_0\left[\frac{2\kappa}{at}ke^{i\theta}+(\frac{a^{2}t^{2}}{\Delta^{2}}+\frac{D}{\Delta})k^{2}e^{-2i\theta}\right].
\label{PminusTW}
\end{align}
\end{subequations}
At finite $\mathbf k$, the polarization selection rules for the band-to-band transitions are still the elliptical ones, but the ellipticity and major axis of the elliptical polarization as functions of $\mathbf k$ are modified by the trigonal warping as well as the effective mass correction term $D \sigma_z k^2$ (c.f. Fig.1(c)), which leads to the modification of the excitonic transition matrix element as well.

With the reduced in-plane rotational symmetry, the excitonic transition selection rules become
\begin{subequations}
\label{TWselectionrules}
\begin{align}
&\left\langle \Psi_{X}\right|\mathbf{\epsilon}_{+}\cdot\hat{\mathbf{p}}\left|0\right\rangle=I'_{s}\delta_{m,0},  \label{TWselectionrule1} \\
&\left\langle \Psi_{X}\right|\mathbf{\epsilon}_{-}\cdot\hat{\mathbf{p}}\left|0\right\rangle=I'_{p}\delta_{m,1}+I'_{d}\delta_{m,-2},  \label{TWselectionrule2}
\end{align}
\end{subequations}
where
\begin{align}
&I'_{s}\sim\sqrt{S}p_0\int_{0}^{2a_{B}^{-1}}
 \rho_{n,0}^{*}(k)[1-(\frac{a^{2}t^{2}}{\Delta^{2}}+\frac{D}{\Delta})k^{2}]kdk, \notag \\
&I'_{p}\sim-\sqrt{S}p_0\int_{0}^{2a_{B}^{-1}}
 \rho_{n,1}^{*}(k)\frac{2\kappa}{at}k^{2}dk,  \notag \\
&I'_{d}\sim-\sqrt{S}p_0\int_{0}^{2a_{B}^{-1}}
 \rho_{n,-2}^{*}(k)(\frac{a^{2}t^{2}}{\Delta^{2}}+\frac{D}{\Delta})k^{3}dk.  \label{dipolestrengthTW}
\end{align}
Here we have neglected the trigonal warping effect on the exciton envelope function $\Phi_{X}(\mathbf{k})$. Without this simplification, Eq. (\ref{TWselectionrule1}) and (\ref{TWselectionrule2}) still hold as they are dictated by the discrete three-fold rotational symmetry, while the quantitative values of $I'_{s}$, $I'_{p}$ and $I'_{d}$ will change.
In Eq. (\ref{TWselectionrule2}), the term $I'_{p}\delta_{m,1}$ means that $p_{+}$ state is now a bright state that can be excited by $\sigma -$ light at K valley. The strength of this transition is proportional to $\kappa$, the degree of trigonal warping. For the -K valley, it is the $p_{-}$ state that becomes bright and can be excited by $\sigma +$ polarized light.

For an order of magnitude estimate, we can write the integral as $\int \rho_{nm}(k) k^l dk \sim k_n^l$, where $k_{n}$ characterises the $\mathbf k$-space spreading of the envelope function~\cite{2D hydrogen k-space wf}. Therefore, the transition strengthes to the $ns$-, $np$- and $nd$-states are respectively
\begin{subequations}
\label{strengthwkn}
\begin{align}
I'_{s} &\sim \sqrt{S} p_0 k_{n}, \label{IsTW} \\
I'_{p} &\sim -\sqrt{S} p_0\frac{2\kappa}{at}k_{n}^{2}, \label{Ip} \\
I'_{d} &\sim -\sqrt{S} p_0(\frac{a^{2}t^{2}}{\Delta^{2}}+\frac{D}{\Delta})k_{n}^{3}, \label{IdTW}
\end{align}
\end{subequations}
where we keep only the leading term of $k_n$. Note that $I'_{d}$ originates from $p_{-}$ in the massive Dirac fermion model (c.f. Eq. (\ref{MDFp-})) which is a second order term in $\mathbf{k}$, while $I'_{s}$ originates from $p_{+}$ which is zeroth order in $\mathbf{k}$. Since $k_n$ is inversely proportional to the Bohr radius which is typically much larger than the lattice constant $a$, $a k_n$ is a small parameter, and the transition matrix element of the $d$ state is in general weak. The effective mass correction to the massive Dirac cones can enhance the transition matrix element of the exciton $d$-state if $D$ is positive (i.e. reducing the electron and hole masses). The transition to the $p$-state is brought in by the trigonal warping, and the transition matrix element is proportional to the degree of warping.

\section{Selection rules for intra-excitonic transitions \label{sectionIntra}}
\begin{figure}
\includegraphics[width=8cm]{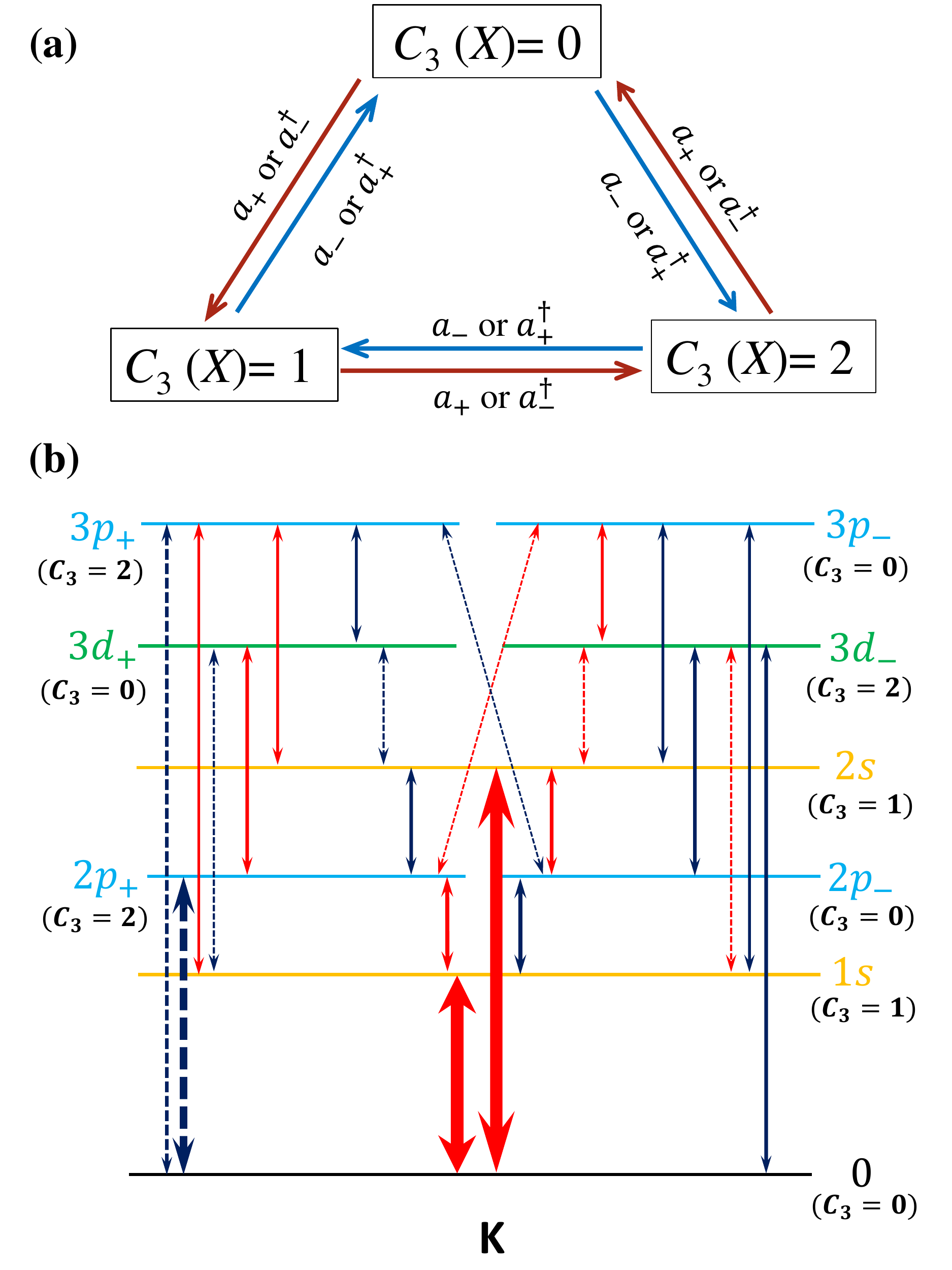}
\caption{(a) Transition selection rules dictated by the three-fold in-plane rotation symmetry of the hexagonal 2D lattices. $C_{3}(X)$ are the quantum numbers associated with the $2 \pi/3$ in-plane rotation for the excitonic states (see text). The red (blue) arrows stand for absorbing a $\sigma +$ ($\sigma -$) polarized photon, or emitting a $\sigma -$ ($\sigma +$) polarized photon. (b) Schematics of the allowed excitonic transition at the K valley of monolayer TMDs. The red (blue) double arrows denote the coupling by $\sigma +$ ($\sigma -$) polarized light. The dash ones are the transitions enabled by trigonal warping effects, and are absent when there is continuous in-plane rotational symmetry. We have dropped nonsecular transitions (i.e. the transition from lower energy states to higher energy ones by emitting photon, or from higher energy ones to lower energy ones by absorbing photon). Thickness of the arrow indicates the strength of the transition (see Table I). The spacing of excitonic states are given according to the first principle calculation in Ref. \onlinecite{X Zhang S Louie TPA 2014}. }
\end{figure}

Having considered the one-photon generation of excitons in the $s$-, $p$- and $d$- states, we now turn to the intra-excitonic optical transition which is of high relevance in the monolayer TMDs because of the large energy separations. Recent first-principle calculations and experiments indicate that the energy spacings between the exciton levels with different $n$ or $|m|$ can range from a few tens to a few hundred meV.~\cite{S. Louie MoS2 specturm calculation PRL2013, X Zhang S Louie TPA 2014} Such large intra-excitonic energy spacings provide new opportunities of probing various intra-excitonic transitions using light sources of infrared frequency range,~\cite{Exciton 1s-2p TMDs nmat2015}
compared to exciton systems in GaAs quantum wells where the intra-excitonic transitions are accessible only by the terahertz lasers due to the much smaller energy splitting.~\cite{Chemla Ultrafast terahertz probes 2003, Broadband THz study of excitonic resonances 2005, Stimulated Terahertz Emission 2006, Transient terahertz spectroscopy of excitons 2009} We show that due to the discrete three-fold rotational symmetry of the system, the intra-excitonic transitions also follow certain selection rules under the excitation by $\sigma \pm$ polarized light, which allows selective accessibility of intra-excitonic transitions.

We first establish the selection rules by symmetry analysis. Use the operator $\hat{C}_{3}$ to denote a $2\pi/3$ in-plane rotation, under which the 2D hexagonal lattice remains unchanged. Consider an exciton state formed at K valley with zero center-of-mass momentum: $\Psi_{X}=\sum_{\mathbf{k}}\Phi_{X}(\mathbf{k})\phi_{c,\mathbf{k}}
(\mathbf{r}_{e})\phi_{v,\mathbf{k}}^{*}(\mathbf{r}_{h})$. Under the $2\pi/3$ in-plane rotation, the Bloch function transforms as
\begin{align}
\hat{C}_{3}\phi_{{j},\mathbf{k}}(\mathbf{r})=e^{-i2(m_{j}+1)\pi/3}\phi_{m_{j},\mathbf{k'}}(\mathbf{r})  \notag
\end{align}
where $j=c, v$ with $m_{c}=0$, $m_{v}=2$ the $d$-orbital quantum number for conduction and valence band at K valley, $\mathbf{k'} \equiv \hat{C}_{3} \mathbf{k}$.
Therefore
\begin{align}
\hat{C}_{3}\Psi_{X}=e^{-i2\pi/3}\sum_{\mathbf{k}}\Phi_{X}(\mathbf{k}) \phi_{c,\mathbf{k'}}(\mathbf{r}_{e})\phi_{v,\mathbf{k'}}^{*} (\mathbf{r}_{h})
\end{align}
And for the envelope function of $s$, $p_{\pm}$ and $d_{\pm}$ states $\Phi_{s}(\mathbf{k})=\Phi_{s}(\mathbf{k}')$, $\Phi_{p\pm}(\mathbf{k})=e^{\mp i2\pi/3}\Phi_{p\pm}(\mathbf{k}')$ and $\Phi_{d\pm}(\mathbf{k})=e^{\pm i2\pi/3}\Phi_{d\pm}(\mathbf{k}')$. Therefore, $\Psi_{X}$ is an eigenstate of $\hat{C}_{3}$,
\begin{align}
\hat{C}_{3}\Psi_{X}=e^{-iC_{3}(X)\cdot2\pi/3}\Psi_{X}, \notag
\end{align}
and for the different orbitals we find the corresponding quantum numbers: $C_{3}(p_{-})=C_{3}(d_{+})=0$, $C_{3}(s)=1$, $C_{3}(p_{+})=C_{3}(d_{-})=2$ and $C_{3}(0)=0$.

We can establish the identity for the optical transition matrix element of absorbing (emitting) a $\sigma \pm$ ($\sigma \mp$) polarized photon:
\begin{eqnarray}
& &\left\langle \Psi_{X'}\right|\mathbf{\epsilon}_{\pm}\cdot\hat{\mathbf{p}}\left|\Psi_{X}\right\rangle \notag \\
&=& \left\langle \Psi_{X'}\right| \hat{C}_3^{-1} \hat{C}_3 \mathbf{\epsilon}_{\pm}\cdot\hat{\mathbf{p}} \hat{C}_3^{-1} \hat{C}_3 \left|\Psi_{X}\right\rangle \notag \\
&=& e^{-i (C_{3}(X) \pm1 -C_{3}(X')) \cdot2\pi/3}  \left\langle \Psi_{X'}\right|\mathbf{\epsilon}_{\pm}\cdot\hat{\mathbf{p}}\left|\Psi_{X}\right\rangle
\end{eqnarray}
A nonzero matrix element $\left\langle \Psi_{X'}\right|\mathbf{\epsilon}_{\pm}\cdot\hat{\mathbf{p}}\left|\Psi_{X}\right\rangle$ therefore requires
\begin{align}
C_{3}(X)  \pm 1 =C_{3}(X') + 3N, \label{selectionrules}
\end{align}
This selection rule has an intuitive meaning from the view of angular momentum conservation: $C_{3}(X)$ is the angular momentum quantum in the intial state, $ \pm 1$ the change of angular momentum quantum by the absorption (emission) of a $\sigma \pm $ ($\sigma \mp$) photon, $C_{3}(X')$ is the angular momentum quantum in the final state, and $3N$ is supplied by the lattice as it only has the 3-fold discrete rotational symmetry. Replacing the initial state by the vaccum which has $C_{3}=0$,  the selection rules for the one-photon generation of excitons discussed in the previous sections are also given by Eq. (\ref{selectionrules}).

Using the massive Dirac fermion model with the trigonal warping effect (Eq. (\ref{2band TW H})), we show below how the transitions allowed by Eq.~(\ref{selectionrules}) emerge, and calculate the strength of the allowed transitions. The intra-excitonic transitions matrix element is
\begin{align}
\left\langle \Psi_{X'}\right|\mathbf{\epsilon}_{\pm}\cdot\hat{\mathbf{p}}\left|\Psi_{X}\right\rangle =\sum_{\mathbf{k}}\Phi_{X'}^{*}\Phi_{X}[p_{cc,\pm}(\mathbf{k})-p_{vv,\pm}(\mathbf{k})] \label{intramatrixelement}
 \end{align}
where $\mathbf{p}_{cc(vv)}(\mathbf{k})=\frac{m_{0}}{\hbar}\langle u_{c(v),\mathbf k}|\nabla_{\mathbf k}\hat{H}_{0}|u_{c(v),\mathbf{k}}\rangle$ are intraband momentum matrix elements and $p_{\pm}(\mathbf{k})=p_{x}(\mathbf{k})\pm ip_{y}(\mathbf{k})$ are given by

\begin{subequations}
\label{intrabandp}
\begin{align}
p_{cc,\pm}(\mathbf{k})=p_{0}\left(\frac{B+D}{2at}+\frac{at}{\Delta}\right)ke^{\pm i\theta}-\frac{3p_{0}\kappa}{\Delta}k^{2}e^{\mp2i\theta}, \label{pcc} \\
p_{vv,\pm}(\mathbf{k})=p_{0}\left(\frac{B-D}{2at}-\frac{at}{\Delta}\right)ke^{\pm i\theta}+\frac{3p_{0}\kappa}{\Delta}k^{2}e^{\mp2i\theta}. \label{pvv}
\end{align}
\end{subequations}

We have
\begin{align}
 \left\langle \Psi_{X'}\right|\mathbf{\epsilon}_{\pm}\cdot\hat{\mathbf{p}}\left|\Psi_{X}\right\rangle =I_{1}\delta_{\Delta m,\pm1}+I_{2}\delta_{\Delta m,\mp2}, \label{exiton-exciton element}
\end{align}
where $\Delta m\equiv m'-m$ is the angular quantum number change in the exciton envelope function, giving exactly the same rules as Eq. (\ref{selectionrules}). In the above, we have dropped terms with $| \Delta m | > 3$. The values are given by
\begin{align}
&I_{1}=2\pi p_0\int\rho_{n'm'}^{*}(k)\rho_{nm}(k)\left(\frac{2at}{\Delta}+\frac{D}{at}\right)k^{2}dk, \notag \\
&I_{2}=-12\pi p_0 \kappa\int\rho_{n'm'}^{*}(k)\rho_{nm}(k)k^{3}dk. \label{intrabanddipole}
\end{align}
The selection rule carried by the second term in Eq. (\ref{exiton-exciton element}) (i.e. $\Delta m=\mp2$) originates from trigonal warping, where the transition strength $I_{2}\propto\kappa$, the degree of the warping.

Fig.2 schematically shows the optical selection rules for both the one-photon generation of excitons and the one-photon intra-excitonic transitions involving exciton energy levels up to $3p$ state in the K valley. The corresponding rules in the -K valley can be obtained by taking the time reversal. The relative strength of the transition matrix elements have been estimated for excitons in monolayer TMDs using different models, as presented in Table I and discussed in the next section.

\begin{table}
\caption{Strength of the transition matrix elements for the one-photon generation of exciton states (upper panel) and for the intra-excitonic transitions (lower panel) in K valley of monolayer WS$_2$. We consider here the absorption of a photon with the specified helicity ($\sigma+$ or $\sigma-$). The columns $\left|\Psi_{X}\right\rangle $ and $\left|\Psi_{X'}\right\rangle$ are the initial and final states respectively, where $\left|0\right\rangle $ denotes the vacuum. The exciton envelope function momentum space spreading $k_n$ is extracted from Ref. [\onlinecite{X Zhang S Louie TPA 2014}] (see main text). The transition matrix elements between the Bloch states are calculated using three different approaches. The magnitudes of one-photon excitonic matrix elements are all normalized by  $|\left\langle \Psi_{1s}\right|\mathbf{\epsilon}_{+}\cdot\hat{\mathbf{p}}\left|0\right\rangle|$ calculated using the corresponding approach, while intra-excitonic transitions are normalized by $|\left\langle \Psi_{2p_{\pm}}\right|\mathbf{\epsilon}_{\pm}\cdot\hat{\mathbf{p}}\left|\Psi_{1s}\right\rangle|$.
''$\star$'' in the mark column denotes that the corresponding transition is enabled by the trigonal warping.}

\begin{ruledtabular}
\label{tableI}
\begin{tabular}{ c | c | c | c | c | c | c  c}
 \multicolumn{2}{c|}{\multirow{2}*{$\Psi_{X'}$}}   & \multicolumn{3}{|c|}{$|\left\langle \Psi_{X'}\right|\mathbf{\epsilon}\cdot\hat{\mathbf{p}}\left|0\right\rangle/\left\langle \Psi_{1s}\right|\mathbf{\epsilon}\cdot\hat{\mathbf{p}}\left|0\right\rangle|$}   & \multirow{2}*{Helicity} &\multirow{2}*{Mark}&\\ \cline{3-5}
\multicolumn{2}{c|}{} & MDF-TW & 3-band TB & DFT &  & \\ \hline
       \multicolumn{2}{c|}{1$s$ }           & 1          &              1                 &              1                &  $\sigma+$ & \\ \hline
          \multicolumn{2}{c|}{2$p_{+}$}      & $3\times10^{-2}$       &  $2\times10^{-2}$  &  $1\times10^{-1}$    &   $\sigma-$ & $\star$  \\ \hline
           \multicolumn{2}{c|}{2$s$ }               & $8\times10^{-1}$        &    $8\times10^{-1}$   &  $8\times10^{-1}$   &   $\sigma+$& \\ \hline
             \multicolumn{2}{c|}{ 3$d_{-}$ }               & $2\times10^{-2}$     &  $2\times10^{-3}$  &  $2\times10^{-4}$    &  $\sigma-$  &\\ \hline
                 \multicolumn{2}{c|}{3$p_{+}$}          & $1\times10^{-3}$      &  $1\times10^{-3}$  &  $1\times10^{-2}$    &  $\sigma-$  & $\star$         \\ \hline    
          \multicolumn{7}{c}{}  \\ \hline
\multirow{2}*{$\Psi_{X}$} & \multirow{2}*{$\Psi_{X'}$}  & \multicolumn{3}{|c|}{$|\left\langle \Psi_{X'}\right|\mathbf{\epsilon}\cdot\hat{\mathbf{p}}\left|\Psi_{X}\right\rangle/\left\langle \Psi_{2p}\right|\mathbf{\epsilon}\cdot\hat{\mathbf{p}}\left|\Psi_{1s}\right\rangle|$}   & \multirow{2}*{Helicity} &\multirow{2}*{Mark}&\\ \cline{3-5}
&  & MDF-TW & 3-band TB & DFT &  & \\  \hline
                             1$s$       &                  2$p_{\pm}$       & 1     & $5\times10^{-1}$  &  1   &  $\sigma\pm$ & \\ \hline
                             1$s$       &                  3$d_{\pm}$   & $2\times10^{-2}$       &  $2\times10^{-2}$   &  $1\times10^{-2}$     &  $\sigma\mp$ & $\star$\\ \hline
                             1$s$       &                  3$p_{\pm}$    & $3\times10^{-1}$       &  $1\times10^{-1}$   &  $2\times10^{-1}$   &  $\sigma\pm$ & \\ \hline
                             2$p_{\pm}$ &             2$s$          & $5\times10^{-1}$        &  $3\times10^{-1}$   &  $5\times10^{-1}$   &  $\sigma\mp$ &  \\ \hline
                 2$p_{\pm}$       &              3$d_{\pm}$        & $1\times10^{-2}$        &  $5\times10^{-3}$   &  $1\times10^{-2}$   &  $\sigma\pm$ & \\ \hline
                 2$p_{\pm}$       &              3$p_{\mp}$     & $5\times10^{-3}$           &  $3\times10^{-3}$   &  $5\times10^{-3}$   &  $\sigma\pm$ & $\star$\\ \hline
                           2$s$        &             3$d_{\pm}$          & $1\times10^{-2}$        &  $8\times10^{-3}$   &  $3\times10^{-3}$   & $\sigma\mp$ &  $\star$\\ \hline
                           2$s$        &             3$p_{\pm}$        & $5\times10^{-1}$          &  $5\times10^{-1}$   &  $5\times10^{-1}$   &  $\sigma\pm$ & \\ \hline
                  3$d_{\pm}$      &             3$p_{\pm}$        & $3\times10^{-1}$         &  $2\times10^{-1}$  & $3\times10^{-1}$   &  $\sigma\mp$ & \\
\end{tabular}
\end{ruledtabular}
\end{table}

\section{Strength of excitonic transitions in monolayer TMDs \label{sectionTMDs}}

In monolayer group VIB  TMDs including MoS$_2$, MoSe$_2$, WS$_2$, and WSe$_2$, the bandgaps are in the visible frequency range, and the energy spacing between the excitonic states are in the  infrared frequency range. These large energy scales allow the optical probe of the various excitonic states for understanding Coulomb interactions in these monolayer semiconductor.~\cite{Heinz binding energy PRL2014, Jie Shan binding energy PRL2014, X Zhang S Louie TPA  2014, XD Cui TPA 2014, Urbaszek TPA and SHG PRL2015, Exciton 1s-2p TMDs nmat2015}
Here we give estimations of the transition strengths in monolayer TMDs, for the allowed excitonic transitions established in the previous sections.

For excitonic states of the Wannier-type wavefunctions, the evaluation of the optical transition matrix elements for the excitonic states requires the knowledge of both the envelop funcitons and the matrix elements of momentum operators between the consituent Bloch states (c.f. Eq. (\ref{matrixelement}) and (\ref{intramatrixelement})). In monolayer TMDs, it has been shown that the distance dependent screening of Coulomb interaction leads to non-hydrogenic envelope functions of excitons \cite{S. Louie MoS2 specturm calculation PRL2013, Heinz binding energy PRL2014, Jie Shan binding energy PRL2014, X Zhang S Louie TPA 2014}. Nevertheless, the radial envelope functions calculated from first principles have qualitatively the same behaviors as the 2D hydrogren wavefunctions, except that the values for $k_n$ are different \cite{X Zhang S Louie TPA 2014}. As shown by Eq.~(\ref{dipolestrengthTW}), (\ref{strengthwkn}), (\ref{intrabanddipole}) and discussions there follows, for an order of magnitude estimation of the transition strength, what matters in the radial envelope functions $\rho_{nm}(k)$ is not its quantitative form, but its momentum space spreading $k_n$. So here in evaluating the excitonic optical matrix elements, we use $k_n$ values extracted from Ref.~[\onlinecite{X Zhang S Louie TPA 2014}]: $k_{0}^{-1}\sim1$ nm, $k_{1}^{-1}\sim1.2$ nm and $k_{2}^{-1}\sim2$ nm for $n=0$, $1$ and $2$ respectively. And the calculated magnitude of 2$p$ state matrix element based on DFT data  (c.f. Table \ref{tableI}) indicates that the oscillator strength of 2$p$ state proportional to $|\left\langle \Psi_{2p_{+}}\right|\mathbf{\epsilon}_{-}\cdot\hat{\mathbf{p}}\left|0\right\rangle|^{2}$ is two orders smaller than that of the 1$s$ state, which is consistent with the calculation in Ref.~[\onlinecite{X Zhang S Louie TPA 2014}]. Thus we expect our approach based on DFT data to give a good order of magnitude estimation to the transition strength.

For evaluating the matrix elements of momentum operators between the Bloch states, we have compared three different approaches. The first is the massive Dirac fermion model with trigonal warping effects (MDF-TW), i. e., $\mathbf k\cdot\mathbf p$ model kept to $k^2$ terms as given in Eq.~(\ref{2band TW H}). The second is a three-band tight-binding model (3-band TB) \cite{Guibin 3bandTB PRB 2014}. Lastly, the parameters in Eq.~(\ref{dipolestrengthTW}), (\ref{strengthwkn}), (\ref{intrabanddipole}) are also determined by fitting the absolute values of the DFT calculated band-to-band momentum matrix element at various $\mathbf k$ points to Eq. (\ref{PplusTW}) and (\ref{PminusTW}). In Table I, we give the strengthes of excitonic optical transition matrix elements, combining the optical matrix elements between the Bloch states with the exciton envelope functions. Through comparing the three different approaches, we find that the  $\mathbf k \cdot\mathbf p$ model and the TB model are not accurate in giving the strength of the optical transition matrix elements. We note that in the 3-band TB model, only the metal atom $d_{z^2}$, $d_{x^2-y^2}$ and $d_{xy}$ orbitals are considered, neglecting other orbitals (e.g., chalcogen atom $p$ orbitals) for simplicity. The $\mathbf k\cdot\mathbf p$ model is further reduced from the TB model by perturbatively eliminating the higher conduction band. The parameters in the models are obtained by fitting the dispersions to the DFT calculated bands in the neighborhood of the K points only \cite{Guibin 3bandTB PRB 2014}. While the dispersions at the band edge are well captured by these models, they are not quantitatively accurate in accounting for quantities such as the interband matrix elements of momentum (which lies in the dependence of the periodic part of the Bloch function on the momentum), due to the oversimplifications in these models. Optical transition strength of excitons are determined by the interband matrix elements of momentum operator. It is thus not surprising that the quantitative values from $\mathbf k\cdot\mathbf p$ model and 3-band TB model are different from the DFT results.

\section{Conclusions}
In summary, we have analyzed the selection rules for the one-photon generation of excitons and for the intra-excitonic optical transitions in massive Dirac cones. We show that the entanglement of the exciton envelope function with the pseudospin texture leads to anomalous selection rules for the one-photon generation of excitons, where the $d$-states become bright, and with opposite valley selection rule from the $s$-states. Such anomalous exciton optical selection rules result from the effect on the interband process by the pseudospin texture, complimentary to the intraband correction (Berry phase correction on exciton binding \cite{Imamoglu p-state Berry splitting 2015, Di Xiao Berry phase and exciton 2015}). The latter effect cannot change the selection rule, but can quantitatively modify the transition dipole strength. Such correction, however, is a small one, which is not explicitly discussed here.

Moreover, in realistic hexagonal 2D lattices, the reduction of the in-plane rotational symmetry, manifested as trigonal warping effects, also modifies the transition selection rules, where $p$-states also becomes bright and have the opposite valley selection rules to the $s$-states. While these selection rules can all be obtained by the analysis of the transformation of exciton envelope functions and the Bloch states under the $2\pi/3$ in-plane rotation, our results show explicitly how the symmetry-allowed transitions missed by the envelope function approximation emerge, when the entanglement of the envelope function with the pseudospin texture is properly accounted.  In monolayer group-VIB TMDs, we have estimated the strength of the one-photon generation of excitonic states, and the transition strength between the excitonic states, by using the first principle calculated envelope function $\mathbf k$-space spreading combined with the optical transition matrix elements between the Bloch states calculated using different approaches.

\section{ACKNOWLEDGMENTS}

The work was supported by the Croucher Foundation (Croucher Innovation Award), the RGC and UGC of HKSAR (HKU705513P, AoE/P-04/08), and the HKU OYRA and ROP.


\begin{thebibliography}{99}
\bibitem{BoxNorm}
The box normalization area $S$ goes away in measurable quantities such as the vacuum Rabi splitting of excitons in a planar cavity which is $\frac{2e}{m_0}\mathbf A_c\cdot\langle\Psi_X|\hat{\mathbf p}|0\rangle$. Here $|\mathbf A_c|\propto 1/\sqrt{S}$ is the average vector potential of a cavity photon.
\end{thebibliography}
\end{document}